\documentclass[11pt]{article}
\usepackage[utf8]{inputenc}
\usepackage{amsfonts}
\usepackage{amsmath}
\usepackage{amssymb}
\usepackage{indentfirst}
\usepackage{graphicx}
\usepackage[colorlinks]{hyperref}
\usepackage{cite}

\usepackage{microtype}
\usepackage[normalem]{ulem}

\numberwithin{equation}{section}
\allowdisplaybreaks

\makeatletter

\begin{document}

\begin{titlepage}
\vspace{3cm}

\baselineskip=24pt

\begin{center}
\textbf{\LARGE{Three-dimensional Newtonian gravity with cosmological constant and torsion}}
\par\end{center}{\LARGE \par}

\begin{center}
	\vspace{1cm}
	\textbf{Patrick Concha}$^{\ast}$,
	\textbf{Evelyn Rodríguez}$^{\ast}$,
	\textbf{Gustavo Rubio}$^{\ast}$,
	\textbf{Paola Yañez}$^{\ddag}$,
	\small
	\\[5mm]
    $^{\ast}$\textit{Departamento de Matemática y Física Aplicadas, }\\
	\textit{ Universidad Católica de la Santísima Concepción, }\\
\textit{ Alonso de Ribera 2850, Concepción, Chile.}
    \\[2mm]
	$^{\ddag}$\textit{Departamento de Física, Universidad de Concepción, }\\ 
    \textit{Casilla,160-C, Concepción, Chile.}
	\\[5mm]
	\footnotesize
	\texttt{patrick.concha@ucsc.cl},
	\texttt{erodriguez@ucsc.cl},
	\texttt{gustavo.rubio@ucsc.cl},
	\texttt{paolayanez@udec.cl},
	\par\end{center}
\vskip 26pt
\begin{abstract}
In this paper we present an alternative cosmological extension of the three-dimensional extended Newtonian Chern-Simons gravity by switching on the torsion. The theory is obtained as a non-relativistic limit of an enhancement and $U(1)$-enlargement of the so-called teleparallel algebra and can be seen as the teleparallel analogue of the Newtonian gravity theory. The infinite-dimensional extension of our result is also explored through the Lie algebra expansion method. An infinite-dimensional torsional Galilean gravity model is presented which in the vanishing cosmological constant limit reproduces the infinite-dimensional extension of the Galilean gravity theory.
\end{abstract}
\end{titlepage}\newpage {} {\baselineskip=12pt \tableofcontents{}}

\section{Introduction}

Newtonian gravity, which is described by the Newton-Cartan (NC) geometry \cite{Cartan1,Cartan2}, has received a growing interest these last years. On one hand, NC geometry has been useful to approach the effective field theory description of the quantum Hall effect \cite{Hoyos:2011ez,Son:2013rqa,Abanov:2014ula,Geracie:2015dea,Gromov:2015fda}. On the other hand, torsional Newton-Cartan (TNC) geometry appears in the study of coupling Newton-Cartan gravity to the Conformal Field theory at the boundary \cite{Christensen:2013lma, Christensen:2013rfa}. Further aplications of the TNC geometry in the context of Lifshitz holography appears in \cite{Hartong:2014oma,Hartong:2014pma,Bergshoeff:2014uea,Bergshoeff:2015ija,Afshar:2015aku}. 

Although NC geometry allows to geometrize the Poisson equation of Newtonian gravity, an action principle for Newtonian gravity requires considering a diverse geometry based on a novel non-relativistic symmetry, which is an extension of the Bargmann algebra \cite{Hansen:2018ofj}. A central extension of such symmetry group, that leaves the Newtonian gravity action invariant, allows us to construct a three-dimensional non-relativistic Chern-Simons (CS) action \cite{Ozdemir:2019orp}. The non-relativistic symmetry is denoted as the extended Newtonian algebra\footnote{also denoted as enhanced Bargmann algebra \cite{Bergshoeff:2020fiz}} and can be obtained as a contraction of the direct sum of the Poincaré and Euclidean algebras. Alternatively, the extended Newtonian symmetry appears as a non-relativistic limit of the coadjoint Poincaré $\oplus\ \mathfrak{u}\left(1\right)^{2}$ algebra \cite{Bergshoeff:2020fiz}. Cosmological extension, Maxwellian generalizations and supersymmetric extension of the extended Newtonian gravity theory were subsequently explored in \cite{Concha:2019dqs,Gomis:2019nih,Concha:2020ebl,Concha:2021jos}.

Three-dimensional CS theories can be seen as toy models since they not only allow us to express gravity as a gauge theory \cite{Achucarro:1987vz,Witten:1988hc,Zanelli:2005sa} but also offers us a more affordable framework to study non-relativistic (super)gravity theories \cite{Bergshoeff:2016lwr,Hartong:2016yrf, Aviles:2018jzw,deAzcarraga:2019mdn,Matulich:2019cdo,Concha:2019lhn,Penafiel:2019czp,Ozdemir:2019tby,Concha:2019mxx,Concha:2020eam}. At the relativistic level, a non-vanishing torsion can be included in a three-dimensional CS gravity through the so-called teleparallel algebra \cite{Caroca:2021njq}. Such theory, which corresponds to a particular case of the Mielke-Baekler gravity model \cite{Mielke:1991nn,Geiller:2020edh}, is characterized by a vanishing curvature and a non-vanishing torsion in which the cosmological constant can be seen as a source for the torsion. At the non-relativistic level, a torsional non-relativistic CS gravity theory can be constructed by applying the non-relativistic limit to the teleparalell $\ \oplus\ \mathfrak{u}\left(1\right)^{2}$ gravity \cite{Concha:2021llq} which in the vanishing cosmological constant limit reproduces the extended Bargmann gravity theory \cite{Bergshoeff:2016lwr}.

In this paper, we accommodate a cosmological constant to the enhanced Bargmann CS gravity by switching on the torsion. To this end we apply the non-relativistic limit to an enhancement of the teleparalell $\ \oplus\ \mathfrak{u}\left(1\right)^{2}$ algebra. The new symmetry is denoted as torsional extended Newtonian algebra and allows us to construct a CS gravity action diverse to the centrally extended post-Newtonian\footnote{also denoted as enhanced Bargmann-Newton-Hooke gravity \cite{Bergshoeff:2019ctr}} one discussed in \cite{Gomis:2019nih}. Analogously to the relativistic teleparallel gravity, the theory is characterized by a non-vanishing spatial torsion and by the vanishing of the curvatures for the non-relativistic spin-connection. Thus, the obtained non-relativistic gravity theory can be seen as the teleparallel analogue of the Newtonian gravity. In the flat limit, the torsion vanish and the theory corresponds to the enhanced Bargmann CS gravity. 

In the second part of this work, we present an infinite-dimensional extension of our results by applying the semigroup expansion (S-expansion) method \cite{Izaurieta:2006zz} to the Poincaré symmetry without considering any enhancement or $U\left(1\right)$-enlargement. The motivation to consider the S-expansion mechanism is twofold. First, the S-expansion method considering $S_{E}$ as the relevant abelian semigroup can be seen as a non-relativistic expansion allowing to derive the respective non-relativistic counterpart of a relativistic symmetry \cite{Gomis:2019nih,Concha:2021jos,Concha:2020tqx}. Secondly, it reproduces the expanded invariant tensor, which are crucial in the construction of a CS action, in terms of the original ones \cite{Izaurieta:2006zz}. Moreover, in our case, the S-expansion procedure offers us a straightforward way to recover the torsional non-relativistic and the torsional extended Newtonian gravities as particular case of an infinite-dimensional torsional Galilean gravity. Finally, the construction of a CS gravity action based on this infinite-dimensional non-relativistic symmetry is presented.

The paper is organized as follows: In section \ref{sec1}, we briefly review the extended Newtonian CS gravity theory defined in three spacetime dimensions. Sections \ref{sec2} and \ref{sec3} contain our main results. In section \ref{sec2}, we first present the torsional extended Newtonian gravity as a non-relativistic limit of the enhanced teleparallel $\ \oplus\ \mathfrak{u}\left(1\right)^{2}$ gravity theory. Section \ref{sec3} is devoted to the construction of an infinite-dimensional extension of our results using the S-expansion procedure. In section \ref{sec4}, we conclude our work with some comments about future developments of our results and the expansion method in the non-relativistic and ultra-relativistic context.


\section{Extended Newtonian gravity theory}\label{sec1}

An action principle for Newtonian gravity requires a particular symmetry algebra which consists of the usual generators of the Bargmann algebra $\{J_{ab},G_{a},H,P_{a},M\}$ \cite{LL,Grigore:1993fz, Bose:1994sj,Duval:2000xr,Jackiw:2000tz,Papageorgiou:2009zc,Andringa:2010it} together with a set of additional generators $\{S_{ab},B_{a},T_{a}\}$ \cite{Hansen:2018ofj}. In three spacetime dimensions, the construction of a proper CS action based on this symmetry requires to introduce two additional central charges $Y$ and $Z$ whose presence ensures a non-degenerate invariant bilinear trace \cite{Ozdemir:2019orp}. The complete set of generators satisfies the extended Newtonian algebra whose commutation relations are given by
\begin{eqnarray}
\left[ J,G_{a}\right] &=&\epsilon _{ab}G_{b}\,, \qquad %
\left[ G_{a},G_{b}\right] =-\epsilon _{ab}S\,, \qquad %
\left[ H,G_{a}\right] =\epsilon _{ab}P_{b}\,,  \notag
\\
\left[ J,P_{a}\right] &=&\epsilon _{ab}P_{b}\,,\qquad \, %
\left[ G_{a},P_{b}\right] =-\epsilon _{ab}M\,,\quad \ \ %
\left[ H,B_{a}\right] =\epsilon _{ab}T_{b}\,,  \notag
\\
\left[ J,B_{a}\right] &=&\epsilon _{ab}B_{b}\,,\qquad %
\left[ G_{a},B_{b}\right] =-\epsilon _{ab}Z\,,\qquad \,  %
\left[ S,G_{a}\right] =\epsilon _{ab}B_{b}\,,
\notag \\
\left[ J,T_{a}\right] &=&\epsilon _{ab}T_{b}\,,\qquad \  %
\left[ G_{a},T_{b}\right] =-\epsilon _{ab}Y\,,\qquad \    %
\left[ S,P_{a}\right] =\epsilon _{ab}T_{b}\,,
\notag \\
\left[ M,G_{a}\right] &=&\epsilon _{ab}T_{b}\,,\qquad \  %
\left[ P_{a},B_{b}\right] =-\epsilon _{ab}Y\,, \label{EN}
\end{eqnarray}
where $a=1,2$, $\epsilon_{ab}=\epsilon_{0ab}$, $\epsilon^{ab}=\epsilon^{0ab}$ and $\epsilon_{ab}\epsilon^{bc}=-\delta_{b}^{\ c}$. Let us note that the extended Bargmann algebra \cite{Bergshoeff:2016lwr,Hartong:2016yrf} appears by setting $B_{a}$, $T_{a}$, $Y$ and $Z$ to zero. The extended Newtonian algebra \eqref{EN} admits the following non-vanishing components of the invariant tensor \cite{Ozdemir:2019orp,Concha:2019dqs}:
\begin{eqnarray}
\langle S S \rangle&=&\langle J Z \rangle=-\beta_0\,, \notag \\
\langle G_a B_b \rangle&=&\beta_0\delta_{ab}\,, \notag \\
\langle M S \rangle&=&\langle H Z\rangle=\langle J Y \rangle=-\beta_1\,, \notag \\
\langle P_a B_b \rangle&=&\langle G_a T_b \rangle=\beta_1 \delta_{ab}\,, \label{IT1}
\end{eqnarray}
where $\beta_0$ and $\beta_1$ are arbitrary constants. Here $\beta_0$ is related to an exotic sector of the extended Newtonian gravity \cite{Concha:2019dqs}. Interestingly, as was shown in \cite{Concha:2019dqs}, the extended Newtonian algebra can also admits the extended Bargmann invariant tensor:
\begin{eqnarray}
\langle J S \rangle&=&-\alpha_0\,,\notag \\ 
\langle G_a G_b \rangle&=&\alpha_0\delta_{ab}\,, \notag\\ 
\langle J M \rangle&=&\langle H S\rangle=-\alpha_1\,, \notag \\ 
\langle G_a P_b \rangle&=&\alpha_1 \delta_{ab}\,. \label{IT2}
\end{eqnarray}
Both families of invariant tensor given by \eqref{IT1} and \eqref{IT2} allow us to construct the most general three-dimensional CS action for the extended Newtonian algebra \cite{Concha:2019dqs}. To this end, let us consider the gauge connection one-form $A$ for the extended Newtonian algebra:
\begin{eqnarray}
A=\omega J+\omega^{a}G_a+\tau H+e^{a}P_{a}+mM+sS+t^{a}T_a+b^{a}B_a+yY+zZ\,. \label{1f}
\end{eqnarray}
Then, the three-dimensional CS action based on the extended Newtonian algebra \eqref{EN} is obtained considering the gauge connection one-form \eqref{1f} and the non-vanishing components of the invariant tensor \eqref{IT1}-\eqref{IT2} into the general expression of the three-dimensional CS action,
\begin{eqnarray}
I_{CS}=\frac{k}{4\pi}\int\langle AdA+\frac{2}{3}A^3\rangle\,,\label{CS}
\end{eqnarray}
where $k$ is the CS level of the theory which is related to the gravitational constant $G$ through $k=1/(4G)$. Thus, the most general extended Newtonian CS gravity action reads, up to boundary terms, as follows \cite{Concha:2019dqs}:
\begin{eqnarray}
I_{\text{Extended-Newtonian}}=\frac{k}{4\pi}\int \mathcal{L}_{\text{Extended-Bargmann}} +\mathcal{L}_{\text{Enhanced-Bargmann}}\,,\label{ENCS}
\end{eqnarray}
where
\begin{eqnarray}
\mathcal{L}_{\text{Extended-Bargmann}}&=&\alpha_0\left[\omega_aR^{a}\left(\omega^{b}\right)-2sR\left(\omega\right)\right] \notag \\
& & +\,2\alpha_1\left[e_aR^{a}\left(\omega^{b}\right)-mR\left(\omega\right)-\tau R\left(s\right) \right]\,,\label{ENH}
\end{eqnarray}
and
\begin{eqnarray}
\mathcal{L}_{\text{Enhanced-Bargmann}}&=&\beta_0\left[b_aR^{a}\left(\omega^{b}\right)+\omega_aR^{a}\left(b^{b}\right) - 2zR\left(\omega\right)-sds \right] \notag \\
& & +2\beta_1\left[e_aR^{a}\left(b^{b}\right)+t_aR^{a}\left(\omega^{b}\right)-yR\left(\omega\right)-\tau R\left(z\right)-mR\left(s\right)\right].\label{eBNH}
\end{eqnarray}
Here, the curvature two-forms are given by
\begin{eqnarray}
R\left(\omega\right)&=&d\omega\,, \notag \\
R\left(s\right)&=&ds + \frac{1}{2} \epsilon^{ac} \omega_a \omega_c \,, \notag \\
R\left(z\right)&=&dz + \epsilon^{ac} \omega_a b_c\,, \notag\\
R^{a}\left(\omega^{b}\right)&=&d\omega^{a}+\epsilon^{ac}\omega\omega_c\,, \notag\\
R^{a}\left(b^{b}\right)&=&db^{a}+\epsilon^{ac}\omega b_c+\epsilon^{ac}s\omega_c\,.\label{curv1}
\end{eqnarray}
The three-dimensional non-relativistic action \eqref{ENCS} describes the most general CS action for the extended Newtonian gravity theory \cite{Concha:2019dqs} which is different from the Newtonian gravity one introduced in \cite{Hansen:2018ofj}. One can notice that the CS action contains two independent sectors given by the Extended-Bargmann gravity action \cite{Bergshoeff:2016lwr,Hartong:2016yrf} and the so-called Enhanced-Bargmann action \cite{Bergshoeff:2020fiz}. In particular, each sector contains an exotic term proportional to $\alpha_0$ and $\beta_0$. Interestingly, the extended-Bargmann gravity theory can be obtained as a non-relativistic limit of the $\mathfrak{iso}\left(2,1\right)\oplus\mathfrak{u}\left(1\right)^{2}$ gravity. On the other hand, the Enhanced-Bargmann gravity theory appears as a non-relativistic limit of the coadjoint Poincaré $\oplus\mathfrak{u}\left(1\right)^{2}$ gravity theory \cite{Bergshoeff:2020fiz}. An alternative procedure to recover the complete extended Newtonian gravity action is the semigroup expansion method \cite{Izaurieta:2006zz}. Indeed, the extended Newtonian algebra can be recovered by expanding the $\mathfrak{iso}\left(2,1\right)$ Lie algebra considering a particular semigroup \cite{Gomis:2019nih}. Such method allows us to obtain the complete set of non-vanishing components of the invariant tensor \eqref{IT1} and \eqref{IT2} in terms of the Poincaré ones.

\section{Newtonian gravity with cosmological constant and torsion}\label{sec2}

The inclusion of a cosmological constant in the extended Newtonian gravity can be done by considering a central extension of the post-Newtonian symmetry \cite{Concha:2019dqs,Gomis:2019nih,Bergshoeff:2020fiz}. The presence of a cosmological constant in non-relativistic gravity was also discussed in \cite{Bacry:1968zf, Aldrovandi:1998im, Gibbons:2003rv, Brugues:2006yd, Alvarez:2007fw, Duval:2011mi, Duval:2016tzi}. Here we present an alternative method to accommodate a cosmological constant to the extended Newtonian gravity by considering a novel non-relativistic symmetry with a non-vanishing spatial torsion. In particular, a torsional extended Newtonian algebra can be obtained by applying the non-relativistic limit to an enhancement and $U\left(1\right)$-enlargement of the so-called teleparallel algebra.

\subsection{The Enhanced teleparallel \texorpdfstring{$\oplus\, \mathfrak{u}\left(1\right)^{2}$}{u1} algebra}
A three-dimensional teleparallel CS formulation of gravity has been recently presented in \cite{Caroca:2021njq}. The teleparallel CS action is gauge invariant under a teleparallel algebra which is spanned by the generators $\{\tilde{J}_{A},\tilde{P}_{A}\}$. In particular, the teleparallel generators satisfy the following non-vanishing commutation relations:
\begin{eqnarray}
\left[\tilde{J}_A,\tilde{J}_B\right]&=&\epsilon_{ABC} \tilde{J}^{C} \,, \notag \\
\left[\tilde{J}_A,\tilde{P}_B\right]&=&\epsilon_{ABC} \tilde{P}^{C} \,, \notag \\
\left[\tilde{P}_A,\tilde{P}_B\right]&=&-\frac{2}{\ell}\epsilon_{ABC} \tilde{P}^{C} \,, \label{telea} 
\end{eqnarray}
where $A,B,C=0,1,2$ and the $\ell$ parameter is related to the cosmological constant $\Lambda$ through $\Lambda \propto - \frac{1}{\ell^2}$.
The teleparallel CS gravity is characterized by a non-vanishing torsion in which the cosmological constant can be seen as a source for the torsion. Then, the geometry is no longer Riemannian but corresponds to the so-called Weizenböck geometry. A teleparallel analogue of the Newtonian gravity can be constructed considering a new non-relativistic symmetry. To this end, let us first consider an enhancement of the teleparallel algebra \eqref{telea} by adding two additional generators $\tilde{S}_{A}$, $\tilde{T}_{A}$. The enhanced teleparallel algebra satisfies \eqref{telea} along with the following commutators:
\begin{eqnarray}
\left[\tilde{J}_A,\tilde{S}_B\right]&=&\epsilon_{ABC} \tilde{S}^{C} \,, \qquad \qquad
\left[\tilde{J}_A,\tilde{T}_B\right]=\epsilon_{ABC} \tilde{T}^{C} \,, \notag \\
\left[\tilde{S}_A,\tilde{P}_B\right]&=&\epsilon_{ABC} \tilde{T}^{C} \,, \qquad \qquad
\left[\tilde{T}_A,\tilde{P}_B\right]=-\frac{2}{\ell}\epsilon_{ABC} \tilde{T}^{C} \,. \label{EnhTel} 
\end{eqnarray}
In the vanishing cosmological constant limit $\ell\rightarrow\infty$ we recover the coadjoint Poincaré algebra \cite{Barducci:2019jhj,Barducci:2020blv,Bergshoeff:2020fiz}. On the other hand, one can notice that the enhanced teleparallel algebra can be written as two copies of the $\mathfrak{iso}\left(2,1\right)$ algebra
\begin{eqnarray}
\left[\tilde{J}_A^{\pm},\tilde{J}_B^{\pm}\right]&=&\epsilon_{ABC} \tilde{J}^{C,\pm} \,, \notag \\
\left[\tilde{J}_A^{\pm},\tilde{P}_B^{\pm}\right]&=&\epsilon_{ABC} \tilde{P}^{C,\pm} \,, \label{2iso}
\end{eqnarray}
by considering the following redefinition of the generators,
\begin{eqnarray}
\tilde{J}_{A}&=&\tilde{J}_{A}^{+}-\tilde{J}_{A}^{-}\,, \qquad \qquad \, \tilde{P}_{A}=-\frac{2}{\ell}\tilde{J}_{A}^{-}\,, \notag \\
\tilde{S}_{A}&=&\tilde{P}_{A}^{+}-\tilde{P}_{A}^{-}\,, \qquad \qquad \tilde{T}_{A}=-\frac{2}{\ell}\tilde{P}_{A}^{-}\,. \label{REDEF1}
\end{eqnarray}
Although one could apply a non-relativistic limit to the enhanced teleparallel algebra \eqref{telea} and \eqref{EnhTel}, we require to consider $\mathfrak{u}\left(1\right)$ generators, $\tilde{Y}_1$ and $\tilde{Y}_2$, in order to obtain a new non-relativistic symmetry diverse to the torsional non-relativistic symmetry introduced recently in \cite{Concha:2021llq}. Furthermore, as we shall see, the presence of $\mathfrak{u}\left(1\right)$ generators would allow to get a non-degenerate invariant tensor in the non-relativistic limit. At the relativistic level, the non-vanishing components of the invariant tensor for the enhanced teleparallel $\oplus\,\mathfrak{u}\left(1\right)^2$ algebra are given by
\begin{eqnarray}
\langle\tilde{J}_{A}\tilde{J}_{B}\rangle&=&\tilde{\alpha_0}\eta_{AB}\,,\qquad \qquad \langle\tilde{J}_{A}\tilde{P}_{B}\rangle=\tilde{\alpha}_{1}\eta_{AB}\,, \qquad \qquad \langle\tilde{P}_{A}\tilde{P}_{B}\rangle=-\frac{2}{\ell}\tilde{\alpha}_{1}\eta_{AB} \,, \notag \\
\langle\tilde{J}_{A}\tilde{S}_{B}\rangle&=&\tilde{\beta_0}\eta_{AB}\,,\qquad \qquad \langle\tilde{P}_{A}\tilde{S}_{B}\rangle=\tilde{\beta}_{1}\eta_{AB}\,, \qquad \qquad \, \langle\tilde{P}_{A}\tilde{T}_{B}\rangle=-\frac{2}{\ell}\tilde{\beta}_{1}\eta_{AB} \,, \notag \\
\langle\tilde{J}_{A}\tilde{T}_{B}\rangle&=&\tilde{\beta}_{1}\eta_{AB}\,, \qquad \qquad \ \, \langle\tilde{Y}_1\tilde{Y}_{1}\rangle=\tilde{\alpha}_{0}\,, \qquad \qquad \qquad \ \langle\tilde{Y}_1\tilde{Y}_{2}\rangle=\tilde{\alpha}_{1}\,, \notag \\
\langle\tilde{Y}_2\tilde{Y}_{2}\rangle&=&-\frac{2}{\ell}\tilde{\alpha}_{1}\,, \label{IT3}
\end{eqnarray}
where $\tilde{\alpha}_{0}$, $\tilde{\alpha}_{1}$, $\tilde{\beta}_{0}$ and $\tilde{\beta}_{1}$ are arbitrary constants. In absence of $\mathfrak{u}\left(1\right)$ generators, the $\tilde{\alpha}$'s are the usual teleparallel gravity constants \cite{Caroca:2021njq}. Let us note that in the flat limit $\ell\rightarrow\infty$ we recover the invariant tensor for the coadjoint Poincaré $\oplus\,\mathfrak{u}\left(1\right)^2$ symmetry.

\subsection{Torsional extended Newtonian algebra}
A non-relativistic version of the enhanced teleparallel $\oplus\,\mathfrak{u}\left(1\right)^2$ algebra appears after considering an Inönü-Wigner contraction of \eqref{telea} and \eqref{EnhTel}. Let us express the relativistic generators as a linear combination of the non-relativistic ones through a dimensionless parameter $\xi$,
\begin{eqnarray}
\tilde{J}_{0} &=&\frac{J}{2}  -\xi ^{4}Z\,,\qquad  \qquad \qquad \ \ \tilde{J}_{a}=\frac{%
\xi }{2}G_{a}-\frac{\xi ^{3}}{2}B_{a}\,,  \notag \\
\tilde{P}_{0} &=&\frac{H}{2}  -\xi ^{4}Y\,,\qquad  \qquad \qquad \ \tilde{P}_{a}=\frac{%
\xi }{2}P_{a}-\frac{\xi ^{3}}{2}T_{a}\,,  \notag \\
\tilde{S}_{0} &=&-\xi ^{2}S-\xi^{4}Z\,,\qquad \qquad \quad \, \tilde{S}_{a}=-\xi G_{a}-\xi ^{3}B_{a}\,,  \notag \\
\tilde{T}_{0} &=&-\xi ^{2}M-\xi^{4}Y\,,\qquad \qquad  \ \  \tilde{T}_{a}=-\xi P_{a}-\xi ^{3}T_{a}\,,  \notag \\
\tilde{Y}_{1}&=& \frac{J}{2}  +\xi ^{4}Z\,, \qquad \qquad \qquad \ \ \tilde{Y}_{2}=\frac{H}{2}  +\xi ^{4}Y\,. \label{NR1}
\end{eqnarray}
Then, in the limit $\xi\rightarrow\infty$, the non-relativistic generators satisfy the commutation relations of the extended Newtonian algebra \eqref{EN} along with:
\begin{eqnarray}
\left[ H,P_{a}\right] &=&-\frac{2}{\ell}\epsilon _{ab}P_{b}\,, \qquad %
\left[ P_{a},P_{b}\right] =\frac{2}{\ell}\epsilon _{ab}M\,, \qquad %
\left[ H,T_{a}\right] =-\frac{2}{\ell}\epsilon _{ab}T_{b}\,,  \notag
\\
\left[ M,P_{a}\right] &=&-\frac{2}{\ell}\epsilon _{ab}T_{b}\,,\qquad \, %
\left[ P_{a},T_{b}\right] =\frac{2}{\ell}\epsilon _{ab}Y\,. \label{TEN}
\end{eqnarray}
The novel non-relativistic symmetry is denoted as Torsional extended Newtonian (TEN) algebra and leads to the torsional non-relativistic algebra introduced recently in \cite{Concha:2021llq} for $B_{a}=T_{a}=Y=Z=0$. As we shall see, the two central charges $Z$ and $Y$ appearing in the TEN algebra will be crucial to define a non-degenerate bilinear form which ensures the proper construction of a CS action. On the other hand, in the vanishing cosmological constant limit $\ell\rightarrow\infty$, we recover the usual extended Newtonian algebra \eqref{EN}. Let us note that the following redefinition of the TEN generators,
\begin{eqnarray}
J&=&J^{+}+J^{-}\,, \qquad \qquad \ \, H=-\frac{2}{\ell}J^{-}\,, \notag \\
G_{a}&=&G^{+}_{a}+G^{-}_{a}\,, \qquad \qquad P_{a}=-\frac{2}{\ell}G^{-}_{a}\,, \notag \\
S&=&S^{+}+S^{-}\,, \qquad \qquad \ M=-\frac{2}{\ell}S^{-}\,, \notag \\
B_{a}&=&B^{+}_{a}+B^{-}_{a}\,, \qquad \qquad \, T_{a}=-\frac{2}{\ell}B^{-}_{a}\,, \notag \\
Z&=&B^{+}+B^{-}\,, \qquad \qquad \ Y=-\frac{2}{\ell}B^{-}\,, \label{REDEF2}
\end{eqnarray}
allows to write the TEN algebra \eqref{EN} and \eqref{TEN} as two copies of the so-called enhanced Nappi-Witten algebra \cite{Bergshoeff:2020fiz,Concha:2020ebl},
\begin{eqnarray}
\left[ J^{\pm},G^{\pm}_{a}\right] &=&\epsilon _{ab}G^{\pm}_{b}\,,\quad \ \left[ G^{\pm}_{a},G^{\pm}_{b}%
\right] =-\epsilon _{ab}S^{\pm}\,,\quad \ \ \left[ S^{\pm},G^{\pm}_{a}\right] =\epsilon
_{ab}B^{\pm}_{b}\,,  \notag \\
\left[ J^{\pm},B^{\pm}_{a}\right] &=&\epsilon _{ab}B^{\pm}_{b}\,,\quad \ \left[ G^{\pm}_{a},B^{\pm}_{b}%
\right] =-\epsilon _{ab}B^{\pm}\,.  \label{eNW}
\end{eqnarray}%
Let us note that the algebra \eqref{eNW} reproduces two copies of the original Nappi-Witten algebra \cite{Nappi:1993ie,Figueroa-OFarrill:1999cmq} when we set $B_{a}^{\pm}=0$. The enhanced version of the Nappi-Witten algebra can be obtained as an IW contraction of the $\mathfrak{iso}(2,1) \oplus \,\mathfrak{u}(1)$ algebra \cite{Bergshoeff:2020fiz} and turned out to be useful to derive diverse Newtonian symmetries through the Semigroup expansion method \cite{Concha:2020ebl,Concha:2021jos}.

\subsection{Newtonian Chern-Simons gravity action with cosmological constant and torsion}

The explicit non-relativistic CS action for the TEN algebra \eqref{EN} and \eqref{TEN} is obtained considering the non-vanishing components of an invariant tensor for the TEN symmetry along with the gauge connection one-form $A=A_{\mu}dx^{\mu}$ in the general expression of the CS action \eqref{CS}. In particular, one can show that the TEN algebra admits the following non-vanishing components of the invariant tensor:
\begin{eqnarray}
\langle S S \rangle&=&\langle J Z \rangle=-\beta_0\,, \notag \\
\langle G_a B_b \rangle&=&\beta_0\delta_{ab}\,, \notag \\
\langle M S \rangle&=&\langle H Z\rangle=\langle J Y \rangle=-\beta_1\,, \notag \\
\langle P_a B_b \rangle&=&\langle G_a T_b \rangle=\beta_1 \delta_{ab}\,, \notag \\
\langle H Y\rangle&=&\langle M M \rangle=\frac{2}{\ell}\beta_1\,, \notag \\
\langle P_{a} T_{b}\rangle &=&-\frac{2}{\ell}\beta_{1}\delta_{ab}\,. \label{IT4}
\end{eqnarray}
which can be obtained from the relativistic ones \eqref{IT3} by applying the limit $\xi\rightarrow\infty$ after considering the contraction of the generators \eqref{NR1} and the rescaling of the relativistic parameters as
\begin{eqnarray}
\tilde{\alpha}_{0}=\tilde{\beta}_{0}=-\beta_0\xi^{4}\,, \qquad \qquad \tilde{\alpha}_{1}=\tilde{\beta}_{1}=-\beta_1\xi^{4}\,. \label{REDEF3}
\end{eqnarray}
One can notice that the exotic sector of the theory proportional to $\beta_0$ is not affected by the inclusion of a cosmological constant and coincides with the extended Newtonian one \eqref{IT1}. It is interesting to note that the TEN algebra also admits the bilinear invariant trace of the torsional non-relativistic algebra introduced in \cite{Concha:2021llq},
\begin{eqnarray}
\left\langle J S\right\rangle &=&-\alpha_{0}\,,  \notag \\
\left\langle G_{a} G_{b}\right\rangle &=&\alpha_{0}\delta _{ab}\,,  \notag \\
\left\langle G_{a} P_{b}\right\rangle &=&\alpha_{1}\delta _{ab}\,,  \notag \\ 
\left\langle H S\right\rangle &=&\left\langle M J\right\rangle =-\alpha_{1}\,,  \notag \\
\left\langle P_{a} P_{b}\right\rangle &=& =-\frac{2\alpha_1}{\ell}\delta _{ab}\,,
\notag \\
\left\langle H M\right\rangle &=&\frac{2\alpha_{1}}{\ell}\,,  \label{IT5}
\end{eqnarray}%
where the relativistic parameters obey $\tilde{\alpha}_{0}=\tilde{\alpha}_{1}=0$ along the following rescaling:
\begin{equation}
\tilde{\beta}_{0}=-\alpha_{0} \xi^{2}\,, \qquad \qquad \tilde{\beta}_{1}=-\alpha_{1} \xi^{2} \,.  \label{alphas}
\end{equation}
Let us note that the components of the invariant tensor proportional to $\alpha$'s are degenerate for the whole TEN algebra although they define non-degenerate invariant trace for the torsional non-relativistic symmetry \cite{Concha:2021llq}. Since each family of invariant tensor appears by applying different rescaling of the relativistic parameters, in the sequel we shall consider the CS construction by exploiting the components given by \eqref{IT4} in order to introduce a torsion in CS Newtonian gravity.

The gauge connection one-form $A$ for the TEN algebra is given by
\begin{equation}
    A=\omega J+\omega^{a} G_{a}+\tau H+e^{a}P_{a}+m M+s S+ t^{a}T_{a}+b^{a}B_{a}+y Y+z Z \,. \label{gc1f}
\end{equation}
The corresponding curvature two-form $F=dA+\frac{1}{2}[A,A]$ reads
\begin{eqnarray}
F&=&R(\omega) J+R^{a}(\omega^{b}) G_{a}+R(\tau) H+R^{a}(e^{b})P_{a}+R(m) M+R(s) S \notag\\
&&+ R^{a}(t^{b})T_{a}+R^{a}(b^{a})B_{a}+R(y) Y+R(z) Z\,,\label{curv2}
\end{eqnarray}
where
\begin{eqnarray}
R(\omega ) &=&d\omega\,,  \notag\\
R(s) &=&ds+\frac{1}{2}\epsilon ^{ac}\omega _{a}\omega _{c}\,, \notag\\
R(z) &=&dz+\epsilon ^{ac}\omega _{a}b_{c}\,, \notag \\
R(\tau ) &=&d\tau \,, \notag \\
R(m) &=&dm+\epsilon ^{ac}\omega _{a}e_{c}-\frac{1}{l}\epsilon ^{ac}e_{a}e_{c}\,, \notag
\\
R(y) &=&dy+\epsilon ^{ac}\omega _{a}t_{c}+\epsilon ^{ac}b_{a}e_{c}-\frac{2}{l%
}\epsilon ^{ac}e_{a}t_{c}\,, \notag \\
R^{a}(\omega ^{b}) &=&d\omega ^{a}+\epsilon ^{ac}\omega \omega _{c}\,, \notag \\
R^{a}(b^{b}) &=&db^{a}+\epsilon ^{ac}\omega b_{c}+\epsilon ^{ac}s\omega _{c}\,, \notag
\\
R^{a}(e^{b}) &=&de^{a}+\epsilon ^{ac}\omega e_{c}+\epsilon ^{ac}\tau \omega
_{c}-\frac{2}{l}\epsilon ^{ac}\tau e_{c}\,, \notag \\
R^{a}(t^{b}) &=&dt^{a}+\epsilon ^{ac}\omega t_{c}+\epsilon
^{ac}se_{c}+\epsilon ^{ac}\tau b_{c}+\epsilon ^{ac}m\omega _{c}-\frac{2}{l}%
\epsilon ^{ac}\tau t_{c}-\frac{2}{l}\epsilon ^{ac}me_{c}\,. \label{ccurv}
\end{eqnarray}
A three-dimensional extended Newtonian CS gravity action with a non-zero torsion based on the TEN algebra \eqref{EN} and \eqref{TEN} can be constructed considering the gauge connection one-form \eqref{gc1f} and the non-vanishing components of the invariant tensor \eqref{IT4} in the general expression of the CS action \eqref{CS}:
\begin{eqnarray}
I_{\text{TEN}} &=&\frac{k}{4\pi}\int \mathcal{L}_{\text{Enhanced-Bargmann}} +\frac{2\beta_{1}}{\ell}\bigg[ \tau R\left(y\right) + y R\left(\tau\right) +m R\left(m\right) -e_{a}R^{a}\left(t^{b}\right) \bigg. \notag \\
&&\bigg. -t_{a}R^{a}\left(e^{b}\right) +\frac{1}{\ell}\epsilon^{ac}\left(me_{a}e_{c}+2\tau t_{a}e_{c}\right) \bigg]\,, \label{TENCS}
\end{eqnarray}%
where $\mathcal{L}_{\text{Enhanced-Bargmann}}$ corresponds to the Enhanced-Bargmann gravity Lagrangian which is given by \eqref{eBNH}. The CS action \eqref{TENCS} is gauge invariant under the whole TEN algebra \eqref{EN} and \eqref{TEN} and can be seen as a cosmological extension of the Enhanced-Bargmann CS gravity theory diverse to the centrally extended post-Newtonian one discussed in \cite{Concha:2019dqs,Gomis:2019nih,Bergshoeff:2020fiz}. In particular, the exotic term proportional to $\beta_0$ is not affected by the inclusion of a cosmological constant. Such exotic term can alternatively be obtained as non-relativistic
limit of the exotic Einstein gravity Lagrangian in the Euclidean and Lorentzian signatures. The cosmological constant contribution appears exclusively along the $\beta_1$ sector. However, the presence of a cosmological constant through the TEN algebra has consequences at the dynamic level. Indeed, requiring the non-degeneracy of the invariant tensor \eqref{IT4} that is $\beta_1\neq0$ and $\beta_{0}\neq-\frac{\ell}{2}\beta_1$, the field equations coming from the CS action \eqref{TENCS} are given by the vanishing of the curvature two-forms \eqref{ccurv}. In particular, on-shell we find
\begin{eqnarray}
T^{a}\left(e^{b}\right)&=&\frac{2}{\ell}\epsilon^{ac}\tau e_{c}\,,\notag \\
T\left(m\right)&=&\frac{1}{\ell}\epsilon^{ac}e_{a}e_{c}\,, \notag \\
T^{a}\left(t^{b}\right)&=&\frac{2}{\ell}\bigg(\epsilon^{ac}\tau t_c + \epsilon^{ac} m e_{c} \bigg) \,, \notag \\
T\left(y\right)&=&\frac{2}{\ell}\epsilon^{ac}e_{a}t_{c}\,. \label{fe}
\end{eqnarray}
where the cosmological constant appears as a source for the spatial torsion $T^{a}\left(e^{b}\right)=de^{a}+\epsilon ^{ac}\omega e_{c}+\epsilon ^{ac}\tau \omega_{c}$ and for the curvatures:
\begin{eqnarray}
T\left(m\right)&=&dm+\epsilon ^{ac}e_{a}\omega _{c}\,, \notag \\
T^{a}\left(t^{b}\right)&=&dt^{a}+\epsilon ^{ac}\omega t_{c}+\epsilon
^{ac}se_{c}+\epsilon ^{ac}\tau b_{c}+\epsilon ^{ac}m\omega _{c}\,, \notag \\
T\left(y\right)&=&dy+\epsilon ^{ac}\omega _{a}t_{c}+\epsilon ^{ac}b_{a}e_{c}\,.
\end{eqnarray}
On the other hand, on-shell we have the vanishing of the curvatures for the spatial and time-like spin-connection $R\left(\omega\right)=0=R^{a}\left(\omega^{b}\right)$. Such particular behavior can be seen as the teleparallel analogue of the Newtonian gravity theory \cite{Hayashi:1979qx,Kawai:1993kvr,deAndrade:1997gka,Sousa:2000bc,DeAndrade:2000sf,Giacomini:2006dr,Caroca:2021njq} which is characterized by a non-Riemannian geometry. Nevertheless, our result is quite diverse to the usual Newton-Cartan gravity theory with torsion \cite{Bergshoeff:2015ija,Bergshoeff:2017dqq,VandenBleeken:2017rij} in which the time component of the torsion is non zero. The torsional Newton-Cartan gravity can be obtained by considering a conformal extension of the Bargmann algebra \cite{Bergshoeff:2014uea}. Here our approach allows us to introduce a cosmological constant in Newtonian gravity by adding a non-vanishing spatial torsion considering a CS formulation. 

Let us note that the field equations for the torsional non-relativistic gravity theory given by \cite{Concha:2021llq}
\begin{equation}
R\left(\omega\right)=0\,, \qquad R\left(s\right)=0\,, \qquad R\left(\tau\right)=0\,, \qquad R\left(m\right)=0\,, \qquad R^{a}\left(\omega^{b}\right)=0\,, \qquad R^{a}\left(e^{b}\right)=0 \,, \notag
\end{equation}
are contained in the TEN gravity model. Such behavior may suggests that there exists an infinite-dimensional non-relativistic symmetry that contains the torsional non-relativistic and the TEN gravity theories as particular cases, which in the vanishing cosmological constant limit reproduces the infinite-dimensional extension of the Galilei algebra introduced in \cite{Hansen:2019vqf} and further discussed in \cite{Gomis:2019fdh,Gomis:2019nih}. In the next section we shall present the construction of a CS gravity action based on an infinite-dimensional torsional Galilean symmetry.

\section{Infinite-dimensional torsional Galilean gravity}\label{sec3}

\subsection{Infinite-dimensional torsional Galilean algebra and semigroup expansion method}

An infinite-dimensional torsional Galilean algebra can be obtained by applying the S-expansion method to the teleparallel algebra \eqref{telea}. Although one could explore its construction by considering a non-relativistic limit, it would require starting from an infinite-dimensional relativistic algebra. Here, we consider a more affordable and efficient approach following the one used in \cite{Gomis:2019nih, Concha:2020tqx,Concha:2021jos}. Before applying the S-expansion to the relativistic teleparallel symmetry we require to consider a particular subspace decomposition of the original algebra \eqref{telea} as follows:
\begin{eqnarray}
V_0&=&\{\tilde{J}_0,\tilde{P}_{0}\}\,,\notag \\
V_1&=&\{\tilde{J}_{a},\tilde{P}_{a}\}\,, 
\end{eqnarray}
where $a=1,2$ is the spatial index. Such subspace decomposition satisfies a $\mathbb{Z}_2$ graded-Lie algebra,
\begin{eqnarray}
[V_0,V_0]\subset V_0\,,\qquad \quad [V_0,V_1]\subset V_1\,, \qquad\quad [V_1,V_1]\subset V_0\,.
\end{eqnarray}
Let $S_{E}^{\left(N\right)}=\left\{ \lambda
_{0},\lambda _{1},\ldots ,\lambda _{N+1}\right\}$ be the relevant semigroup whose elements satisfy the following multiplication law:
\begin{equation}
\lambda _{\alpha }\lambda _{\beta }=\left\{ 
\begin{array}{lcl}
\lambda _{\alpha +\beta }\,\,\,\, & \mathrm{if}\,\,\,\,\alpha +\beta \leq
N+1\,, &  \\ 
\lambda _{N+1}\,\,\, & \mathrm{if}\,\,\,\,\alpha +\beta >N+1\,, & 
\end{array}%
\right.   \label{se}
\end{equation}%
Here $\lambda_{N+1}=0_S$ represents the zero element of the semigroup which satisfies $0_S \lambda_i=0_S$ for $i=0,\ldots,N+1$. Let us consider now a subset decomposition of the semigroup $S_{E}^{\left(N\right)}=S_{0}\cup S_{1}$ with
\begin{eqnarray}
S_{0} &=&\left\{ \lambda _{2m},\ \text{with }m=0,\ldots ,\left[ \frac{N}{2}%
\right] \right\} \cup \{\lambda _{N+1}\}\,, \notag \\
S_{1} &=&\left\{ \lambda _{2m+1},\ \text{with }m=0,\ldots ,\left[ \frac{N+1}{%
2}\right] \right\} \cup \{\lambda _{N+1}\}\,,
\end{eqnarray}%
where $[\ldots ]$ denotes the integer part. Such subset decomposition is said to be resonant since it satisfies the same algebraic structure than the subspace decomposition,
\begin{equation}
    S_0\cdot S_0\subset S_0\,,\qquad \quad \ S_0\cdot S_1\subset S_1\,,\qquad \quad \ S_1\cdot S_1\subset S_0\,.\label{semidecomp}
\end{equation}
Then, considering the definitions of \cite{Izaurieta:2006zz}, one obtain a resonant expanded Lie algebra $\mathcal{G}_{R}=W_{0}\oplus W_{1}$ with
\begin{equation}
W_{0}=S_{0}\times V_{0}\,, \qquad \qquad W_{1}=S_{1}\times V_{1}\,.
\end{equation}
After extracting a $0_S$-reduced subalgebra we get an infinite-dimensional expanded algebra whose generators are related to the original ones as follows:
\begin{eqnarray}
J^{(m)} &=&\lambda _{2m}\tilde{J}_{0}\,, \qquad \qquad \,
G_{a}^{(m)} =\lambda _{2m+1}\tilde{J}_{a}\,,  \notag \\
H^{(m)} &=&\lambda _{2m}\tilde{P}_{0}\,,  \qquad \qquad
P_{a}^{(m)} =\lambda _{2m+1}\tilde{P}_{a}\,.  \label{texp}
\end{eqnarray}%
Here the $0_S$-reduction condition implies that $0_S T_A=0$ allowing us to abelianize a large sector of the resonant expanded algebra. In particular, the expanded generators \eqref{texp} satisfy the following non-vanishing commutation relations:
\begin{eqnarray}
\lbrack J^{(m)},G_{a}^{(n)}] &=&\epsilon _{ab}G_{b}^{(m+n)}\,, \qquad \qquad 
\lbrack G_{a}^{(m)},G_{b}^{(n)}] =-\epsilon _{ab}J^{(m+n+1)},  \nonumber \\
\lbrack J^{(m)},P_{a}^{(n)}] &=&\epsilon _{ab}P_{b}^{(m+n)}\,, \qquad \qquad \lbrack G_{a}^{(m)},P_{b}^{(n)}] =-\epsilon _{ab}H^{(m+n+1)},  \nonumber \\
\lbrack H^{(m)},G_{a}^{(n)}] &=&\epsilon _{ab}P_{b}^{(m+n)}\,, \qquad \qquad \lbrack H^{(m)},P_{a}^{(n)}] =-\frac{2}{\ell}\epsilon _{ab}P_{b}^{(m+n)}, 
\nonumber \\
\lbrack P_{a}^{(m)},P_{b}^{(n)}] &=&\frac{2}{\ell}\epsilon _{ab}H^{(m+n+1)}.
\label{ITG}
\end{eqnarray}%
This algebra leads, in the vanishing cosmological constant limit $\ell\rightarrow\infty$, to the infinite-dimensional extension of the Galilean algebra discussed in \cite{Hansen:2019vqf,Gomis:2019fdh,Gomis:2019nih}. Then, the algebra \eqref{ITG} can be seen as an infinite-dimensional torsional Galilean ($\mathfrak{tG}^{\left(N\right)}$) algebra. The non-relativistic algebra \eqref{ITG} corresponds then to a cosmological extension of the infinite-dimensional Galilean algebra diverse to the infinite-dimensional graded algebra obtained in \cite{Gomis:2019nih}. Interestingly, for $N=2$ we recover the torsional non-relativistic algebra presented recently in \cite{Concha:2021llq}. On the other hand, for $N=4$ we get the TEN algebra previously obtained as a non-relativistic limit of the enhanced teleparallel $\oplus\mathfrak{u}\left(1\right)^{2}$ algebra. Although one can construct a CS gravity action for the $N=3$ case, one can show that the bilinear invariant trace is degenerate, making some gauge fields not determined by the field equations, thus reducing the number of dynamical fields. The degeneracy can be avoided for even values of $N$ assuring a kinematical term for each gauge field. A new finite non-relativistic symmetry with non-degenerate invariant tensor appears then for $N=6$.

Let us remark that the infinite dimensional torsional Galilean algebra can be rewritten as two copies of an infinite-dimensional extension of the Nappi-Witten algebra \cite{Bergshoeff:2020fiz}, which we have denoted as $\mathfrak{nw}^{\left(N\right)}$:
\begin{eqnarray}
\lbrack J^{(m)\pm},G_{a}^{(n)\pm}] &=&\epsilon _{ab}G_{b}^{(m+n)\pm}\,,  \nonumber \\
\lbrack G_{a}^{(m)\pm},G_{b}^{(n)\pm}] &=&-\epsilon _{ab}J^{(m+n+1)\pm}\,.
\label{INW}
\end{eqnarray}%
The two copies of $\mathfrak{nw}^{\left(N\right)}$ appear after considering the following redefinition of the $\mathfrak{tG}^{\left(N\right)}$ generators
\begin{eqnarray}
G_{a}^{(m)}&=&G^{(m)+}_a + G^{(m)-}_a\,, \qquad \qquad \ \  P^{(m)}_a=-\frac{2}{\ell}G^{(m)-}_a\,, \notag \\
J^{(m)}&=&J^{(m)+}+J^{(m)-}\,, \qquad \qquad \ \ \ \,  H^{(m)}=-\frac{2}{\ell}J^{(m)-} \,. \label{REDEF4}
\end{eqnarray}
In particular, $\mathfrak{nw}^{\left(2\right)}$ and $\mathfrak{nw}^{\left(4\right)}$ are the usual Nappi-Witten \cite{Nappi:1993ie,Figueroa-OFarrill:1999cmq} and enhanced Nappi-Witten algebras, respectively \cite{Bergshoeff:2020fiz,Concha:2020ebl}.


\subsection{Non-relativistic Chern-Simons actions}

The construction of a CS gravity action based on the infinite-dimensional torsional Galilean symmetry \eqref{ITG} requires to consider a gauge connection one-form,
\begin{equation}
    A=\sum_{m=0}^{\left[\frac{N}{2}\right]}\left(\omega^{(m)}J^{(m)}+\tau^{(m)}H^{(m)}\right)+\sum_{m=0}^{\left[\frac{N+1}{2}\right]}\left(\omega^{a(m)}G_{a}^{(m)}+e^{a(m)}P_{a}^{(m)}\right)\,, \label{infA}
\end{equation}
and the non-vanishing components of the invariant tensor. In particular, one can show that the $\mathfrak{tG}^{\left(N\right)}$ algebra admits the following invariant tensor:
\begin{eqnarray}
\left\langle J^{(m)}J^{(n)}\right\rangle  &=&-\sigma_{m+n}\,,\qquad \qquad \quad \left\langle G_{a}^{(m)}P_{b}^{(n)}\right\rangle  =\gamma _{m+n+1}\,\delta
_{ab},  \nonumber \\
\left\langle G_{a}^{(m)}G_{b}^{(n)}\right\rangle  &=&\sigma_{m+n+1}\delta
_{ab}\,, \qquad \quad \ \, \left\langle P^{(m)}P^{(n)}\right\rangle  =\frac{2}{\ell}\gamma _{m+n}\,, 
\nonumber \\
\left\langle J^{(m)}P^{(n)}\right\rangle  &=&-\gamma _{m+n}\,, \qquad \qquad \quad  \left\langle P_{a}^{(m)}P_{b}^{(n)}\right\rangle  =-\frac{2}{\ell}\gamma
_{m+n+1}\,\delta _{ab},  \label{IIT}
\end{eqnarray}%
where the $\sigma$'s and $\gamma$'s are related to the relativistic coupling constants $\tilde{\alpha}_{0}$ and $\tilde{\beta}_{1}$ of the teleparallel CS gravity theory \cite{Caroca:2021njq} through the semigroup elements as follows
\begin{eqnarray}
\sigma_{m+n} &=&\lambda _{2m+2n}\tilde{\alpha}_{0}\,,  \nonumber \\
\gamma_{m+n} &=&\lambda _{2m+2n}\tilde{\alpha}_{1}\,.  
\end{eqnarray}
In particular, $\sigma_{1}$ and $\gamma_{1}$ correspond to the respective non-relativistic constants $\alpha_0$ and $\alpha_1$ of the invariant tensor of the torsional non-relativistic symmetry \eqref{IT5}. On the other hand, $\sigma_{2}$ and $\gamma_{2}$ are the respective TEN coupling constants $\beta_{0}$ and $\beta_{1}$ appearing in \eqref{IT4}. Let us note that $\sigma_{0}$ and $\gamma_{0}$ are related to trivial torsional and Galilean contributions which we shall omit in the construction of the CS action.  Then, considering the gauge connection one-form \eqref{infA} for the $\mathfrak{tG}^{\left(N\right)}$ algebra and the non-vanishing components of the invariant tensor \eqref{IIT} in the general expression of the three-dimensional CS action \eqref{CS}, we get

\begin{eqnarray}
I_{\mathfrak{tG}^{\left(N\right)}}&=&\frac{k}{4\pi}\sum_{i,m,n,l=1}^{[N/2]}\sigma_{i}\int\left(-\omega^{(m)}d\omega^{(n)}\delta_{m+n}^{i}+\omega_{a}^{(m)}d\omega^{a(n)}\delta_{m+n+1}^{i}-\epsilon^{ac}\omega^{(m)}\omega_{a}^{(n)}\omega_{c}^{(l)}\delta_{m+n+l+1}^{i} \right) \notag \\
& &+\frac{k}{2\pi}\sum_{i,m,n,l=1}^{[N/2]}\gamma_{i}\int\left[ -R\left(\omega^{(m)}\right)\tau^{(n)}\delta_{m+n}^{i}+R^{a}\left(\omega^{b(m)}\right)e_{a}^{(n)}\delta_{m+n+1}^{i} \right. \notag \\
& &\left. +\frac{1}{\ell}R\left(\tau^{(m)}\right)\tau^{(n)}\delta_{m+n}^{i}+\frac{1}{\ell}R^{a}\left(e^{b(m)}\right)e_{a}^{(n)}\delta_{m+n+1}^{i}+\frac{1}{\ell^2}\epsilon^{ac}\tau^{(m)}e_{a}^{(n)}e_{c}^{(l)}\delta_{m+n+l+1}^{i} \right] \,, \notag \\ \label{ITGCS}
\end{eqnarray}
where
\begin{eqnarray}
R\left(\omega^{(m)}\right)&=&d\omega^{(m)}+\sum_{n,l=1}^{[N/2]}\frac{1}{2}\epsilon^{ac}\omega_{a}^{(n)}\omega_{c}^{(l)}\delta_{n+l+1}^{m}\,, \notag \\
R^{a}\left(\omega^{b(m)}\right)&=&d\omega^{a(m)}+\sum_{n,l=1}^{[N/2]}\epsilon^{ac}\omega^{(n)}\omega_{c}^{(l)}\delta_{n+l}^{m}\,, \notag \\
R\left(\tau^{(m)}\right)&=&d\tau^{(m)}+\sum_{n,l=1}^{[N/2]}\epsilon^{ac}\left(\omega_{a}^{(n)}e_{c}^{(l)}\delta_{n+l+1}^{m}-\frac{1}{\ell}e_{a}^{(n)}e_{c}^{(l)}\delta_{n+l+1}^{m}\right)\,, \notag \\
R^{a}\left(e^{b(m)}\right)&=&de^{a(m)}+\sum_{n,l=1}^{[N/2]}\epsilon^{ac}\left(\omega^{(n)}e_{c}^{(l)}\delta_{n+l}^{m}+\tau^{(n)}\omega_{c}^{(l)}\delta_{n+l}^{m}-\frac{2}{\ell}\tau^{(n)}e_{c}^{(l)}\delta_{n+l}^{m}\right)\,, \label{infcurv}
\end{eqnarray}
are the respective curvature two-forms for the $\mathfrak{tG}^{\left(N\right)}$ algebra. The CS action \eqref{ITGCS} contains two sectors, one proportional to $\gamma_i$ and a second one proportional to $\sigma_i$, which corresponds to an exotic sector. One can see that each sector is gauge invariant under the $\mathfrak{tG}^{\left(N \right)}$ algebra. Interestingly, considering only the non-degenerate cases given by even values of $N$, one can rewrite the CS action \eqref{ITGCS} as follows:
\begin{eqnarray}
I_{\mathfrak{tG}^{\left(N\right)}}&=&I_{\text{TNR}}+I_{\text{TEN}}+\sum_{i=3}^{N/2}I_{\mathfrak{tG}^{\left(2i\right)}}\,, \label{ndITGCS}
\end{eqnarray}
where $I_{\text{TNR}}$ corresponds to the torsional non-relativistic CS action \cite{Concha:2021llq}:
\begin{eqnarray}
I_{\text{TNR}} &=&\frac{k}{4\pi}\int \alpha_{0}\left[ \omega _{a}R^{a}(\omega
^{b})-2sR\left( \omega \right) \right] +\alpha_{1}\left[
2e_{a}R^{a}(\omega ^{b})-2mR(\omega )-2\tau R(s)\right. \notag \\
&&-\left.\frac{2}{\ell}e_{a}R^{a}\left( e^{b}\right)+\frac{2}{\ell}mR\left( \tau \right)+\frac{2}{\ell}\tau R\left(m\right)+\frac{2}{\ell^2}\tau\epsilon^{ac}e_{a}e_{c}\right]  \,  \label{torCS}
\end{eqnarray}%
with $R^{a}\left(\omega^{b}\right)$, $R\left(\omega\right)$, $R\left(s\right)$, $R^{a}\left(e^{b}\right)$, $R\left(\tau\right)$ and $R\left(m\right)$ being defined in \eqref{ccurv}. On the other hand, $I_{\text{TEN}}$ reproduces the torsional extended Newtonian CS action \eqref{TENCS} introduced in the previous section. Let us note that the components of the invariant tensor proportional to $\sigma_{i-1}$ and $\gamma_{i-1}$, although they are non-degenerate for the $\mathfrak{tG}^{\left(2i-2\right)}$ algebra, they are degenerate for the $\mathfrak{tG}^{\left(2i\right)}$ algebra. The degeneracy can be avoided either considering the complete set of invariant tensor proportional to $\sigma_\alpha$ and $\gamma_\alpha$ for $\alpha=1,\cdots,i$ or taking into account only the components along $\sigma_{i}$ and $\gamma_{i}$. In particular, the non-degeneracy of the invariant tensor for the $\mathfrak{tG}^{\left(2i\right)}$ algebra \eqref{IIT} requires only
\begin{eqnarray}
\gamma_{i}\neq 0\,, \qquad \text{and} \qquad \sigma_{i}\neq -\frac{\ell}{2}\gamma_{i}\,,
\end{eqnarray}
without imposing conditions on the other constants. In such case, the equations of motion of the theory are given by the vanishing of curvature two-forms \eqref{infcurv}. On shell, we have that the spatial and time components of the infinite torsion two-forms satisfy
\begin{eqnarray}
T^{a}\left(e^{b(m)}\right)&=& \sum_{n,l=1}^{N/2}\frac{2}{\ell}\epsilon^{ac}\tau^{(n)}e_{c}^{(l)}\delta_{n+l}^{m}\,, \notag \\
T\left(\tau^{(m)}\right)&=&\sum_{n,l=1}^{N/2}\frac{1}{\ell}\epsilon^{ac}e_{a}^{(n)}e_{c}^{(l)}\delta_{n+l+1}^{m}\,,
\end{eqnarray}
where
\begin{eqnarray}
T^{a}\left(e^{b(m)}\right)&=&de^{a(m)}+\sum_{n,l=1}^{N/2}\epsilon^{ac}\left(\omega^{(n)}e_{c}^{(l)}\delta_{n+l}^{m}+\tau^{(n)}\omega_{c}^{(l)}\delta_{n+l}^{m}\right) \,, \notag \\
T\left(\tau^{(m)}\right)&=&d\tau^{(m)}+\sum_{n,l=1}^{N/2}\epsilon^{ac}\omega_{a}^{(n)}e_{c}^{(l)}\delta_{n+l+1}^{m} \,. \label{InfTor}
\end{eqnarray}
Here the cosmological constant can be seen as a source for the torsion two-forms \eqref{InfTor} describing the infinite-dimensional non-relativistic analogue of the teleparallel CS gravity model \cite{Caroca:2021njq}. Moreover, the curvatures for the spatial and time-like infinite-dimensional extension of the spin-connection vanish $R\left(\omega^{(m)}\right)=0=R^{a}\left(\omega^{b(m)}\right)$. Interestingly, in the vanishing cosmological constant limit $\ell\rightarrow\infty$ the non-relativistic CS gravity based on the $\mathfrak{tG}^{\left(2i\right)}$ algebra reduces to the infinite-dimensional extension of the Galilean CS gravity \cite{Gomis:2019nih} with a vanishing torsion.

\section{Conclusions}\label{sec4}

In this work we presented an alternative way to introduce a cosmological constant to the three-dimensional Newtonian gravity, by considering the non-relativistic limit of the enhanced teleparallel $\ \oplus\ \mathfrak{u}\left(1\right)^{2}$ gravity theory. We showed that the new non-relativistic symmetry, denoted as torsional extended Newtonian algebra, can be written as two copies of the so-called enhanced Nappi-Witten one \cite{Bergshoeff:2020fiz,Concha:2020ebl} allowing us to construct a cosmological extension of the enhanced Bargmann gravity action \cite{Bergshoeff:2020fiz} diverse to the centrally extended post-Newtonian one \cite{Concha:2019dqs,Gomis:2019nih}. The presented theory is characterized by a non-vanishing spatial torsion, in which the cosmological constant can be interpreted as a source for the torsion. It is important to clarify that the torsional Newtonian gravity theory obtained here is different to the Newton-Cartan theory with torsion \cite{Bergshoeff:2014uea,Bergshoeff:2015ija,Bergshoeff:2017dqq,VandenBleeken:2017rij} in which the time component of the torsion is non zero. Indeed, our procedure allows us to construct a gauge-invariant non-relativistic CS gravity action which can be seen as the teleparallel analogue of the Newtonian gravity. It would be interesting to study how our methodology could be extended in order to make contact with torsional Newton-Cartan formalism from a CS point of view.

We have also generalized our results by presenting an infinite-dimensional torsional Galilean symmetry which is obtained by expanding the Poincaré algebra with an infinite semigroup $S_{E}^{\left(N\right)}$. The infinite-dimensional non-relativistic structure contains our torsional extended Newtonian algebra and the torsional non-relativistic algebra presented recently in \cite{Concha:2021llq} as particular (finite) sub-cases for $N=4$ and $N=2$, respectively. The corresponding CS gravity action based on the infinite-dimensional torsional Galilean symmetry was also presented, which in the vanishing cosmological constant limit reproduces the infinite-dimensional extension of Galilean gravity discussed in \cite{Gomis:2019nih}. 

The Lie algebra expansion method has received a growing interest due to its applications not only in the non-relativistic and ultra-relativistic contexts \cite{deAzcarraga:2019mdn,Bergshoeff:2019ctr,Romano:2019ulw,Concha:2019lhn,Penafiel:2019czp,Ozdemir:2019tby,Gomis:2019nih,Concha:2019mxx,Concha:2020eam,Fontanella:2020eje,Gomis:2022spp} but also in approaching asymptotic symmetries \cite{Caroca:2017onr,Caroca:2018obf,Caroca:2019dds}, higher-spin gravity \cite{Caroca:2017izc}, supergravity models \cite{Izaurieta:2006aj,Fierro:2014lka, deAzcarraga:2014jpa,Concha:2014tca,Concha:2019icz}, among others. On the other hand, the construction of non-relativistic (super)gravity actions in higher spacetime dimensions remains a difficult task since the non-relativistic limit cannot be naively applied. The Lie algebra expansion method could be useful to overcome such difficulties and thus to obtain Galiean or Carrollian (super)algebras along their extended versions in arbitrary spacetime dimensions. However, the construction of an action based on these symmetries would require a completely different geometric approach [work in progress].


\section*{Acknowledgements}

The authors would like to thank N. Merino for interesting discussion and useful comments. This work was funded by the National Agency for Research and Development ANID - PAI grant No. 77190078 (P.C.), ANID - SIA grant No. SA77210097 (E.R.) and FONDECYT grants No. 1211077 (P.C.), 11220328 (P.C.) and 11220486 (E.R.). This work was supported by the Research project Code DIREG$\_$09/2020 (P.C.) of the Universidad Católica de la Santísima Concepción, Chile. P.C., E.R. and G.R. would like to thank to the Dirección de Investigación and Vice-rectoría de Investigación of the Universidad Católica de la Santísima Concepción, Chile, for their constant support.

\appendix


\bibliographystyle{fullsort.bst}
 
\bibliography{Newtonian_gravity_and_torsion}

\end{document}